\documentclass[aps,pre,floats, twocolumn,showpacs,superscriptaddress]{revtex4}

\usepackage{graphicx,epsfig}% Include figure files
\usepackage{times}
\usepackage{graphics,dcolumn,bm,epic, eepic,fleqn,float}
\usepackage{amssymb,amsmath,multirow,rotate,color}

\newcommand{\alessio}[1]{\textcolor{black}{#1}}
\usepackage[normalem]{ulem}

\begin{document}
%\topmargin 0.1cm
%\color{black}
\title{The Evolutionary Vaccination Dilemma in Complex Networks}

\author{Alessio Cardillo}
\affiliation{Departamento de F\'{\i}sica de la Materia Condensada, University of Zaragoza, Zaragoza 50009, Spain}
\affiliation{Institute for Biocomputation and Physics of Complex Systems (BIFI), University of Zaragoza, Zaragoza 50018, Spain}

\author{Catalina Reyes-Su\'arez}
\affiliation{Departamento de F\'{\i}sica, Universidad Pedag\'ogica y Tecnol\'ogica de Colombia, Tunja, Colombia}

\author{Fernando Naranjo}
\affiliation{Departamento de F\'{\i}sica, Universidad Pedag\'ogica y Tecnol\'ogica de Colombia, Tunja, Colombia}

\author{Jes{\'u}s G{\'o}mez-Garde\~{n}es}
%\email{gardenes@gmail.com}
\affiliation{Departamento de F\'{\i}sica de la Materia Condensada, University of Zaragoza, Zaragoza 50009, Spain}
\affiliation{Institute for Biocomputation and Physics of Complex Systems (BIFI), University of Zaragoza, Zaragoza 50018, Spain}

\date{\today}

\begin{abstract}
In this work we analyze the evolution of voluntary vaccination in networked populations by entangling the spreading dynamics of an influenza-like disease with an evolutionary framework taking place at the end of each influenza season so that individuals take or not the vaccine \alessio{upon} their previous experience. Our framework thus put in competition two well-known dynamical properties of scale-free networks: the fast propagation of diseases and the promotion of cooperative behaviors. Our results show that when vaccine is perfect scale-free networks enhance the vaccination behavior with respect to random graphs with homogeneous connectivity patterns. However, when imperfection appears we find a cross-over effect so that the number of infected (vaccinated) individuals increases (decreases) with respect to homogeneous networks, thus showing up the competition between the aforementioned properties of scale-free graphs.
\end{abstract}

\pacs{89.75.Fb, 05.70.Fh}

\maketitle

\section{Introduction}

The advent of network science \cite{rev:albert,rev:newman} has provided, an important set of computational and statistical physics tools for describing the problem \alessio{of} epidemic spreading by incorporating the realistic interaction patterns of the constituents of social and technological systems \cite{rev:bocc}. Classical approaches to epidemiology \cite{book1,book2} rely on the use of the theory of phase transitions and critical phenomena, so to unveil the onset and the macroscopic impact of epidemic outbreaks. Recently, these techniques have been pervasively adapted to study a variety of critical phenomena on top of networks \cite{rev:doro}.

The main contribution of the former line of research to epidemiology has been the development of a generalized mean-field framework in which general patterns of interactions can be included. In particular, it was shown \cite{vespi,vespi2,may,moreno,marc,gomez2} that for scale-free networks [in which the probability distribution of having a node with $k$ neighbors follows a power-law, $P(k)\sim k^{-\alpha}$] the epidemic onset was anticipated as compared to substrates with more regular (or homogeneous) connectivity patterns. Moreover, when $\alpha<3$ (as most of social and technological networks show \cite{sex1,amaral-2000}) and for large enough (thermodynamic limit) systems, the epidemic onset vanishes, meaning that even a very small fraction of infected elements with small infective power can spread a disease to a macroscopic part of the population by a sequence of contagions between neighbors of the network, as it happens in human contacts \cite{sex2,sex3,sex4,sex5}.

Apart from the theoretical value of the above finding, its direct implications \alessio{on} public health campaigns and the security of technological networks such as the Internet demand a deeper understanding about the influence that diverse contact patterns have on disease dynamics, its co-evolution \cite{Gross,Guerra} and the design of new algorithms for immunization and vaccination policies. Typically, these studies aim at identifying the most efficient way for reducing the impact of an epidemic by the vaccination/immunization of the minimal number of nodes. To this aim, different methods to identify the most important nodes to be immunized have been proposed \cite{imm1,imm2,cover,imm3}.

The former works concern the immunization of technological networks since in social contexts vaccination is typically voluntary. Thus, the study of the immunization of a population demands to include the ways vaccination and risky behaviors compete and spread across individuals. To this aim, one \alessio{may} consider game theory  to formulate a social dilemma in terms of the benefits associated to each of the behaviors: vaccination or not. Within this framework individuals act rationally, i.e., by choosing their strategy after an evaluation of their potential benefits. This evaluation is done by considering their perception of the risk to contract the disease. For well-mixed populations recent results show \cite{bauch1,bauch4,vardavas,vardavas2,bauch2,bauch3} that voluntary vaccination is not efficient to reach efficient immunization. However, this kind of approach was generalized to networks \cite{NJP} unveiling an enhancement of voluntary vaccination.

The former game theoretical approach considers that agents aim at maximizing their own benefits. However, the decisions of individuals can evolve in time depending on the epidemic incidence observed in the population. In this framework agents are prone to adopt the strategies that are expected to perform better based on the information available. This evolutionary avenue has been recently adopted to the vaccination dilemma. A first evolutionary avenue is presented in \cite{bhattarcharyya,bauch2012,donofrio2012} where both disease transmission and vaccinating behavior evolve in time simultaneously. The evolution for the fraction of vaccinated individuals is driven by the difference of payoffs between vaccinated and non-vaccinated agents (as in the case of the well-known replicator equation of evolutionary games \cite{gintis,nowak}), being the latter determined by the epidemic incidence at that time. A second evolutionary approach is proposed in \cite{wells2013}. In this case, inspired on seasonal influenza, the number of vaccinated individuals remains constant during the duration of the influenza season. After each season, individuals evaluate the payoffs based on the incidence of the disease in the last season and decide whether to vaccinate or not for the next seasons. 

Here we take a similar avenue to that of \cite{wells2013} regarding the dynamical setup and the motivation: the vaccination for seasonal influeza. However, the way in which payoffs are constructed and the way individuals choose their strategy  follows the typical setup of evolutionary games \cite{gintis,nowak}. This setup, originally presented in \cite{prsb} for the vaccination dilemma, considers that individuals are assigned a payoff that is solely based on the personal experience during the last season. In addition, the strategic choice is based on the imitation of those individuals with better payoffs. Thus, we do not consider that individuals follow a rational derivation of the payoffs associated to vaccination and risky behavior based on the information  available. This allow us to connect the vaccination dilemma with other studies on the evolutionary game dynamics of social dilemmas \cite{gintis,nowak}. 

In the recent years, the study of the evolutionary game dynamics of social dilemmas on structured populations \cite{szabo,anxorev,our} have shown that cooperation (here related to vaccination) is favored when the interactions among individuals take the form of scale-free networks \cite{santos,campillo}. Inspired in this result, in this work we explore the spread of vaccination behavior across networks with homogeneous and heterogeneous (scale-free) connectivity patterns. Our results show that when vaccine is perfect scale-free networks enhance the vaccination behavior with respect to homogeneous graphs, thus reducing the impact of the disease on the population. However, when vaccine is imperfect we find a cross-over effect, and homogeneous networks outperform scale-free \alessio{ones}. This latter scenario reveals an interesting  competition between the rapid spread of both diseases and cooperative behaviors in scale-free graphs.

\section{The Model}

As introduced above, to incoporate the competition between disease spreading and evolutionary dynamics on top of a network we entangle these two dynamical frameworks by producing an iterative sequence of a two-stage process. In both stages the interaction pattern among individuals is described by a complex network (\alessio{keeping} the same network for both the dynamical setups). This network is given by an ($N\times N$) Adjacency matrix $A_{ij}$ so that when two individuals interact $A_{ij}=1$, \alessio{whereas $A_{ij}=0$ otherwise}. In this way, the number $k_i$ of neighbors (contacts) of a given node, say $i$, is given by $k_i=\sum_{j=1}^N A_{ij}$.

\begin{figure}[t!]
\centering
\includegraphics[width=3.450in]{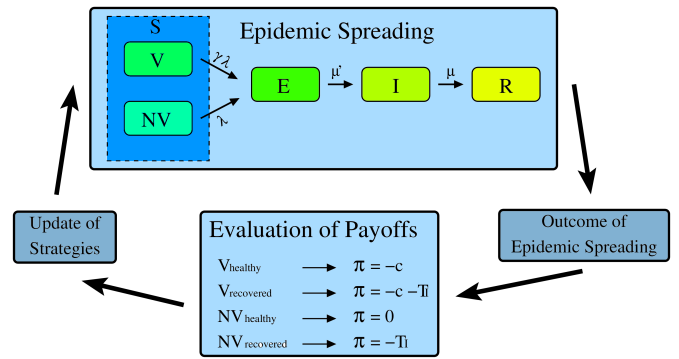}
\caption{(color online). Resuming sequence of the evolutionary picture of our model. The top box descibes the epidemic spreading process. The bottom one, instead, displays the payoffs accumulated by the agents according to their \emph{strategy}. Arrows denote the causal order of the evolutionary process.}
\label{fig1}
\end{figure}

In this work we will consider two of the most paradigmatic network models: Erd\H{o}s-R\'enyi (ER) graphs \cite{ER} and Barab\'asi-Albert (BA) networks \cite{BA}. The former class of graphs are described by a Poisson degree distribution $P(k)$, so that most of the nodes have a connectivity close to the mean value $\langle k\rangle$. On the other hand,  BA networks display a power-law degree distribution of the form, $P(k)\sim k^{-3}$, thus incorporating the scale-free (SF) property of real-world networks.
The implementation of our dynamical setup aims at revealing the differences between the heterogeneous degree pattern displayed in SF and the rather homogeneous structure of ER graphs. 
 {\color{black} To this aim, for both ER and SF networks, the average connectivity of the nodes is set to $\langle k\rangle=6$.}
Below we introduce the rules governing the two-stage dynamics that it is also sketched in Fig. \ref{fig1}.

\subsection{Disease spreading}

The first of the stages of our dynamical setup is based on the evolution of a Susceptible-Exposed-Infected-Recovered (SEIR) model \cite{book1,book2}.  This model captures the dynamics of influenza-type infections. Susceptible nodes have not been infected and are healthy. They catch the disease via direct contact with Exposed neighbors at a rate $\lambda$. Exposed nodes are supposed to carry the virus although they still do not display symptoms of the disease, thus these individuals are highly infectious during this incubation period. Exposed nodes \alessio{become} Infected with some rate $\mu^{\prime}$ which typically is the inverse time of the incubation period of the disease. Infected nodes, on the other hand, although still carrying the virus are here assumed not to be infectious. In particular, we consider that during this period they remain isolated from the rest of the population. Finally, Infected nodes pass to the Recovered state with rate $\mu$ that is the inverse duration time of the convalescence period. 

With the above rules we consider that each node $i$ interacts simultaneously with its $k_i$ neighbors per unit time. Thus, for a network described by the Adjacency matrix $A_{ij}$ the effective probability that a Susceptible node $i$ gets the disease per unit time is given by:
\begin{equation}
P^{i}_{S \rightarrow E}=1-(1-\lambda)^{\sum_{j=1}^N A_{ij}x_j}\;,
\end{equation}
where $x_j=1$ when node $j$ is Exposed and $x_j=0$ otherwise. Here, in order to mimic the transmission of ordinary influenza we have set $\mu^{\prime}=0.33$, since the time elapsed between exposure to the virus and development of symptoms is two to three days. In addition we take $\mu=0.2$ since the symptoms of uncomplicated influenza illness resolve after a period of $3$ to $7$ days, so that the average permanence in the Infected state is $\mu^{-1}=5$ days.

The addition of vaccinated individuals to the formulation of our SEIR model implies that initially there is subset of Susceptible individuals (representing a fraction $N_V$ of the total population) that are less prone to catch the disease than non-vaccinated Susceptible ones. In particular, we consider that a vaccinated individual is infected during a single contact with an Exposed one at a rate $\lambda\cdot\gamma$, where $\gamma\in[0,1]$ is a parameter that \alessio{modulates} the quality of the vaccine, being perfect when $\gamma=0$ and useless for $\gamma=1$. In this way, the probability that a vaccinated individual $i$ is infected per unit time reads:
\begin{equation}
P^{i}_{S \rightarrow E}=1-(1-\gamma\cdot\lambda)^{\sum_{j=1}^N A_{ij}x_j}\;.
\end{equation}

Once the values of the epidemic parameters $\mu$ and $\mu^{\prime}$, the quality $\gamma$ of the vaccine and the fraction $N_V$ of vaccinated individuals are set, we leave $\lambda$ as the relevant control parameter of the SEIR model. In addition, the relevant order parameter of the dynamics is the fraction $R$ of nodes that got infected once the epidemic process dies out, so that the macroscopic behavior is captured \alessio{by} the curve $R(\lambda)$. For a given value of $\lambda$ one starts from an initial state in which a small fraction (here the $5\%$) of the population is set as Exposed. Then the SEIR dynamics is iterated until no individuals remain \alessio{either} as Exposed or Infected. 

\subsection{Evolutionary Dynamics}

Once the SEIR dynamics dies out we consider that the seasonal influenza period has passed. Before the next SEIR dynamics starts, individuals evaluate whether to vaccinate or not for the next season. At this point evolutionary dynamics takes place by assigning to each of the individuals a payoff $\pi_i$ ($i=1,...,N$) that depends on their experience accumulated during the last SEIR propagation. As shown in Fig. \ref{fig1}, there are four possibilities:
\begin{itemize}
\item[\em (i)] Vaccinated individuals that remain healthy during the last season have payoff $\pi=-c$ (where $c$ is a cost associated to the vaccine). 
\item[\em (ii)] Vaccinated individuals that were infected during the last season have payoff $\pi=-c-T_{I}$ (where $T_{I}$ is the time units that the individual remain in the Infected state).
\item[\em (iii)] Individuals that did not vaccinated and remain healthy during the season have payoff $\pi=0$.  
\item [\em (iv)] Non-vaccinated individuals that were infected are assigned a payoff $\pi=-T_{I}$.
\end{itemize}

The cost $c$ associated to the vaccination is related to different issues such as the time spent to get vaccinated (via Public Health Services)  or the probability that the vaccine causes \alessio{side} effects. To illustrate the vaccination dilemma let us show a very simple situation of a susceptible agent $i$ in contact with an exposed agent. In this situation the expected payoff of $i$ when having taken the vaccine is: $\pi^{exp}_{V}=-(1-\gamma\lambda)c-\gamma\lambda(c+1/\mu)$ (here we assume that $T_I\simeq1/\mu$). On the other hand, if agent $i$ has adopted a risky behavior, its expected payoff turns into: $\pi^{exp}_{NV}=-\lambda/\mu$. Thus, in this single pairwise encounter, the rational choice is not to take the vaccine for any costs $c>\lambda(1-\gamma)/\mu$. This simple situation clearly reveals the Vaccination Dilemma. However, in a networked population the situation is rather more complex and, more importantly, here we assume that individuals are not fully rational and, instead of deciding their behavior on expectations, they evolve their strategies based on their previous experience.

Evolutionary dynamics provides the framework to implement the dynamical evolution of strategies. In particular, as it is usually done in evolutionary social dilemmas on networks, each individual, say $i$, chooses at random one of its first neighbors, say $j$, and compares their payoffs $\pi_i$ and $\pi_j$ respectively. Then, the probability that agent $i$ takes the strategy of $j$, $s_j$, for the next season increases with their payoff difference, $(\pi_j-\pi_i)$. One of the most used frameworks to calculate this probability is that of the Fermi-like rule \cite{fermi1,fermi2}, in which the probability that the strategy of the neighbor $j$ is adopted by $i$ reads:
\begin{equation}
P_{s_j\rightarrow s_i}=\frac{1}{1+\exp\left[-\beta(\pi_j-\pi_i)\right]}\;,
\end{equation}
where $\beta$ is a parameter that allows to \alessio{span} between random ($\beta\ll 1$) and strong selection ($\beta\gg 1$). Here we adopted $\beta=1$ and checked that our results are quite robust under changes of $\beta$.
The update of strategies takes place simultaneously for all the agents. \alessio{Once} the new strategies are 
taken, the payoffs are set to zero and the SEIR dynamics starts again with a new fraction $N_V$ of vaccinated susceptible individuals. 

\begin{figure}[t!]
\centering
\includegraphics[width=3.3in]{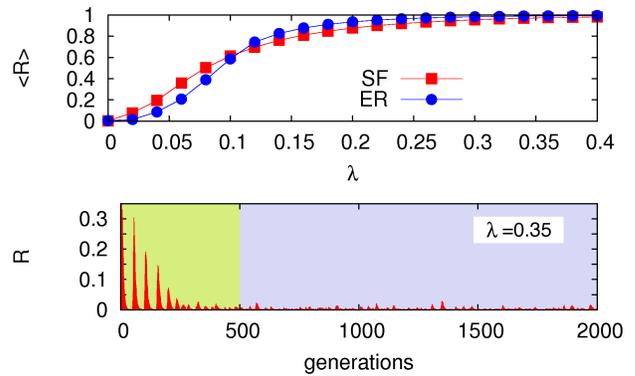}
\caption{(color online). The top panel shows the epidemic diagram $\langle R\rangle(\lambda)$ for ER and SF networks when vaccination is not allowed. The bottom panel shows the evolution of the \alessio{fraction} of Recovered individuals, $R$, with the generations of the evolutionary dynamics. The network is SF and the rate of infection per contact is $\lambda=0.35$, whereas vaccination is perfect $\gamma=0$ and it has a cost $c=0.1$.}
\label{fig2}
\end{figure}

Finally, let us note that we iterate the sequence of the two-stage process (SEIR dynamics and Evolutionary Dynamics) for a number of steps (generations) \alessio{large enough to reach a steady} state for the relevant observables: the average fraction of recovered, $\langle R\rangle$, and  vaccinated individuals, $\langle N_V\rangle$. {\color{black} In addition, at the beginning of each generation we randomly assign the individuals that are vaccinated (so that they constitute the $25\%$ of the population) and those that are initially set as Exposed (reaching the $5\%$ of the total population). It is worth mentioning that in real cases a small fraction of the population gain permanent immunity from the exposure to the virus in the last generation. In our case we do not consider such inherited immunity to the new strain}.

\begin{figure*}[t!]
\centering
\includegraphics[width=6.3in]{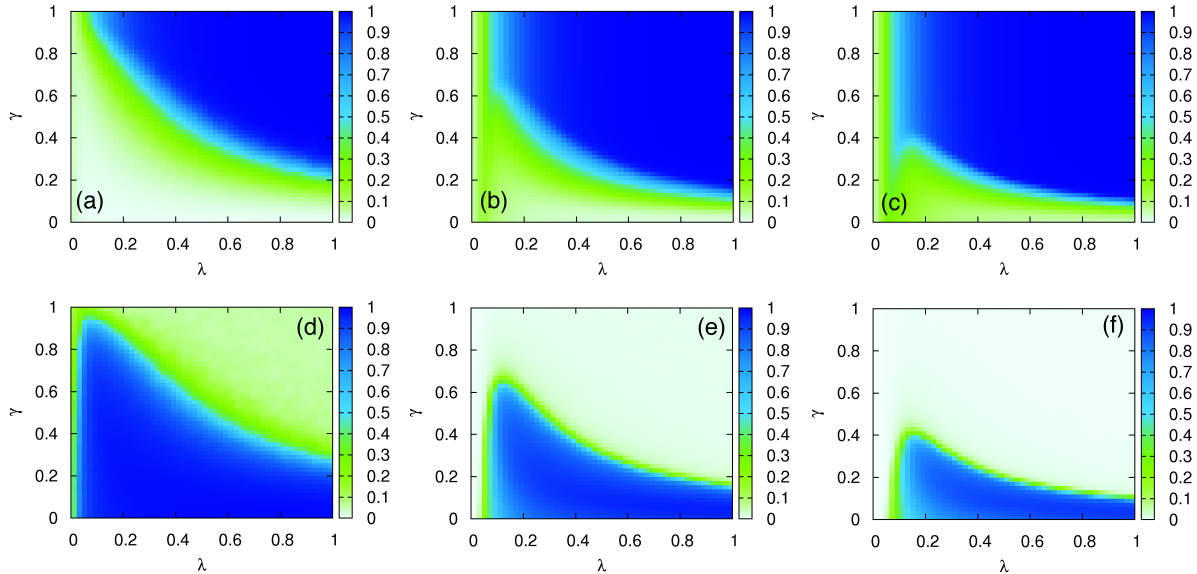}
\caption{(color online). The contour plots show the average fraction of Recovered $\langle R\rangle$ (top) and Vaccinated $\langle N_V \rangle$ (bottom) individuals as a function of the infection rate $\lambda$ and the vaccine quality $\gamma$ for SF networks. From left to right the panels correspond to different vaccination costs: $c=0.1$ [panels (a) and (d)], $c=0.5$ [panels (b) and (e)] and $c=1.0$ [panels (c) and (f)]. As the cost increases we note that the overall fraction of vaccinated individuals decreases while that of Recovered nodes increases. Interestingly when $c=0.1$ there is a range of low $\gamma$ values ($\gamma<0.1$) for which the epidemic threshold disappears and the disease cannot spread for any value of $\lambda$.}
\label{fig5}
\end{figure*}

\section{Results}

We start our discussion by briefly reporting the behavior of the SEIR model without vaccinated individuals. In the top panel of Fig. \ref{fig2} we show the average fraction $\langle R\rangle$ of recovered individuals at the end of the SEIR dynamics as a function of the rate of infection per contact, $\lambda$, for ER and SF networks of $N=1000$. From this figure it becomes clear that SF networks accelerates the onset $\lambda_c$ of the epidemic regime as compared to ER graphs.

Let us now focus on the case \alessio{of} SF networks to evaluate the impact that voluntary vaccination (under an evolutionary framework) has on the immunization of the system. In the bottom panel of Fig. \ref{fig2} we show the evolution of the fraction of recovered individuals $R$ for a sequence of $2000$ generations. The rate of infection used in this simulation is set to $\lambda=0.35$ which, as the top panel shows, corresponds to a situation in which almost all the population has been infected $\langle R \rangle \simeq 1$ when no vaccination is allowed. Instead, when individuals can decide whether to take the vaccination (under the aforementioned evolutionary rules) we show that the epidemic phase does not appear ($R\simeq 0$) since the population has evolutionarily adopted the vaccination strategy. 

Remarkably, the transient regime (lasting around $500$ generations) shows an interesting pattern of rise-and-falls for the number of recovered individuals $R$. This behavior points out that, before vaccination prevails, the population displays an oscillating behavior between vaccination and risky behavior. Obviously, when many people vaccinate (falls in R) the epidemic falls but vaccinated individuals are tempted not to take the vaccine due to the higher benefits of risky individuals. This leads to a progressive increase of the infections (\alessio{denoted} by the increase of $R$) that reverse the balance of benefits between risky and vaccinated individuals. This rise-and-fall behavior together with the significative duration of this transient regime reveal the importance of risk perception in voluntary vaccination.

\subsection{Macroscopic behavior of vaccine taking in SF networks}

Now we analyze the behavior after the transient regime. To this aim we compute the average fraction of vaccinated $\langle N_V\rangle$ and Recovered $\langle R\rangle$ individual in the steady state as a function of $\lambda$ and the quality $\gamma$ of the vaccine. {\color{black} For each couple of values ($\lambda$, $\gamma$) we have run $100$ simulations (each of them comprising $2000$ generations).}
In Fig. \ref{fig5} we report these functions for several vaccine costs $c$ in SF networks. In particular, the panels in the top show the diagrams $\langle R\rangle(\lambda,\gamma)$ and those in the bottom show $\langle N_V\rangle(\lambda,\gamma)$. From left to right the panels correspond to the following vaccine costs: $c=0.1$, $0.5$ and $1.0$. 

Let us focus on those diagrams corresponding to $c=0.1$ [panels (a) and (d)]. The function $R(\lambda,\gamma)$ shows that for values of $\gamma\in[0,0.1]$ (roughly perfect vaccination) the epidemic threshold disappears since $\langle R\rangle\simeq 0$ for all the values of $\lambda$. In its turn, we note from panel (d) that for this latter region the fraction of vaccinated individuals is roughly $\langle N_V\rangle\simeq 1$ except for very low values of $\lambda$ for which the disease cannot spread even when no immunization is present. If we increase further the value of $\gamma$ we recover the epidemic onset $\lambda_c$ whose values decreases as the vaccine get worse, i.e., as $\gamma$ increases. In addition, the vaccination behavior decreases so that for a given value of $\gamma$ the advantage provided by vaccines is not useful anymore for $\lambda>\lambda_c$. Obviously, for $\gamma=1$ we recover the usual diagram $R(\lambda)$, shown in the top panel of Fig. \ref{fig1}, for SF networks since the vaccine provides no advantage and, as shown in panel (d), almost no individual in the network holds the vaccination strategy \alessio{giving} $\langle N_V\rangle\simeq 0$ for all $\lambda$ values.

As we increase the cost of the vaccine to $c=0.5$ [panels (b) and (e)] and $c=1.0$ [panels (c) and (f)] we observe that the overall \alessio{fraction} of Recovered (Vaccinated) individuals increases (decreases). Remarkably, the maximum value of $\gamma$ for which there is no epidemic threshold decreases with $c$ and for $c=1.0$ we cannot appreciate this effect. It is interesting to note that the usual epidemic diagram of SF networks without immunization is recovered for lower values of $\gamma$. For instance, in panel (b) we note that for $\gamma>0.6$ the curve $R(\lambda)$ does not change whereas from panel (e) we note that, within this region, individuals do not vaccinate anymore ($\langle N_V\rangle =0$).

\subsection{SF versus ER networks: The importance of vaccine quality}

Having reported the macroscopic behavior in SF networks as concerns the influence of the vaccine quality and its cost, we now focus on the dependence on the networked substrate in which both the disease and the vaccination strategies spread. To this aim, we compare the behavior in SF and ER networks in order to measure the role of degree heterogeneity on the vaccination behavior. {\color{black} Importantly, we have considered SF networks as obtained from the Barab\'asi-Albert model \cite{BA} after a complete randomization preserving the degree sequence of the nodes. In this way, we obtain SF networks with $P(k)\sim k^{-3}$ without any kind of degree-degree correlations that could influence the dynamical behavior. In addition, we have increased the size of the networks considered (in order to fully exploit the heterogeneous property of SF networks) to $N=10^4$ nodes.}

\begin{figure}[t!]
\centering
\includegraphics[width=3.45in]{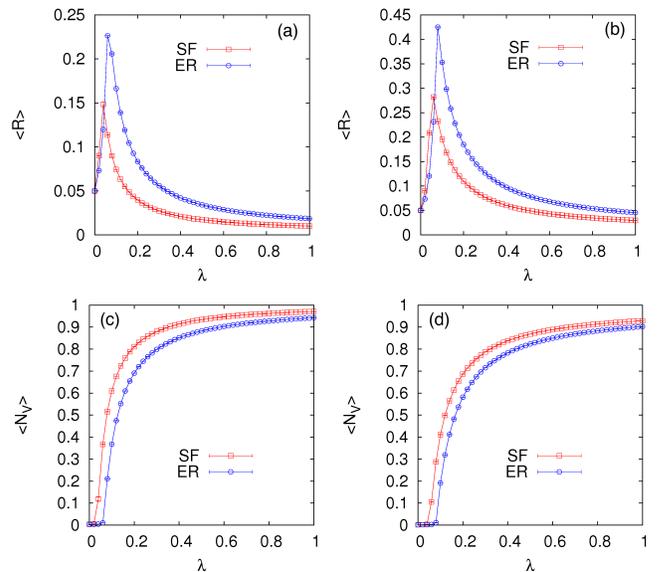}
\caption{(color online). Epidemic $\langle R\rangle(\lambda)$ (top panels) and Vaccination $\langle N_V\rangle(\lambda)$ (bottom panels) diagrams for ER and SF networks ($N=10^4$, $\langle k\rangle=6$) when the vaccine is perfect ($\gamma=0$). The cost associated to the vaccine are $c=0.5$ (left panels) and $c=1.0$ (right panels).}
\label{fig3}
\end{figure}

We first explore the case of perfect vaccination, $\gamma=0$. In Fig. \ref{fig3} we show the diagrams $\langle R \rangle (\lambda)$ (top) and $\langle N_V\rangle(\lambda)$ (bottom) for two different vaccination costs: $c=0.5$ [panels (a) and (c)] and $c=1.0$ [panels (b) and (d)]. In these panels we also show the standard deviations around the average values reported. From the panels we observe that SF networks outperform ER graphs since the overall average number of recovered (vaccinated) individuals is smaller (higher) in SF networks. In particular, the epidemic diagrams $\langle R \rangle (\lambda)$ display a clear peak around the respective epidemic thresholds, $\lambda_c$, of the original (without vaccination) graphs. Up to this point $\lambda<\lambda_c$, the epidemic does not spread and thus vaccination behavior is not observed either as shown in the diagrams $\langle N_V \rangle (\lambda)$. The peak thus point out that the risk is so small that vaccination behavior do not show up and \alessio{leading} to a burst of infections which reaches higher values in ER graphs. This result seems counterintuitive, since from the \alessio{literature on epidemic on epidemics on networks}, SF graphs are always more prone to the spread of diseases than ER ones. Furthermore, from the diagrams $\langle N_V \rangle (\lambda)$ we note that the vaccination onset starts earlier for SF graphs, as their natural epidemic threshold are smaller than that of ER \alessio{ones}.

For values of $\lambda$ above the natural epidemic threshold, the number of Recovered nodes decreases dramatically in both networks. Here, the risk of infection becomes larger and individuals start to adopt the vaccination strategy as diagrams $\langle N_V \rangle (\lambda)$ in panels (c) and (d) show. However, vaccination behavior spreads easier in SF networks than in ER graphs and it is quite remarkable that, for this regime, the number of Recovered nodes in ER graphs is always (for any value of $\lambda$) higher than in SF networks. Thus cooperative behavior, by taking the vaccine, spreads better in SF networks, in agreement with those studies about cooperation and social dilemmas in complex networks \cite{santos,campillo}.

\begin{figure}[t!]
\centering
\includegraphics[width=3.45in]{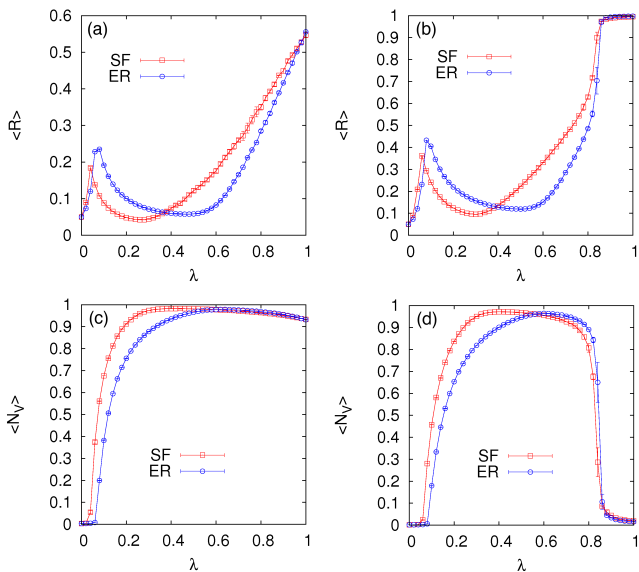}
\caption{(color online).Epidemic $\langle R\rangle(\lambda)$ (top panels) and Vaccination $\langle N_V\rangle(\lambda)$ (bottom panels) diagrams for ER and SF networks ($N=10^4$, $\langle k\rangle=6$) when the vaccine is not perfect ($\gamma=0.12$). The cost associated to the vaccine are $c=0.5$ (left panels) and $c=1.0$ (right panels). The imperfection of the vaccine causes two crossovers, one for $\langle R\rangle$ and the other one for $\langle N_V\rangle$, between the performance of SF networks and ER graphs.}
\label{fig4}
\end{figure}

In Fig. \ref{fig4} we explore the scenario of imperfect vaccination considering $\gamma=0.12$. This regime shows the competition between two well-known effects: the aforementioned prevalence of cooperative behaviors in SF networks (with respect to ER graphs) and their weakness to the spread of diseases (again with respect to ER graphs). This competition appears as a crossover between the behavior of both $\langle R\rangle(\lambda)$ and $\langle N_V\rangle(\lambda)$ in SF and ER networks. In panels (a) and (b) we show that the curves $\langle R\rangle(\lambda)$  (after the peak close to the natural epidemic thresholds of both networks) cross at some $\lambda^*$ values, \alessio{which decreases with the cost of the vaccine $c$}. Panels (c) and (d) show also a crossover behavior for $\langle N_V\rangle (\lambda)$, which appears with some delay with respect to that occurring at $\lambda^*$ for $\langle R\rangle(\lambda)$. {\color{black} Note that this crossover is well defined since the standard deviations around the average values $\langle R\rangle$ and $\langle N_V\rangle$ are extremely low.}

The behavior for $\lambda<\lambda^*$ shows \alessio{the same trend} as for the perfect vaccination case. SF networks outperform ER graphs showing a larger number of vaccinated individuals and a smaller number of infections. However, \alessio{for the imperfect vaccine ($\gamma>0$) the growth of $\lambda$ affects both non-vaccinated and vaccinated individuals. Under such conditions, the virus finds in the SF networks a better backbone to propagate}. In this way, panels (a) and (b) show that the failure of vaccination starts to become evident in SF networks at $\lambda^*$. The smaller benefits provided by the imperfection of the vaccine cause that the number of vaccinated individuals starts to decrease after $\lambda^*$. Being larger the number of infections \alessio{due to} the imperfect vaccine in SF networks, as shown for $\lambda>\lambda^*$, the fall of vaccinated individuals occurs in SF networks at smaller values of $\lambda$ than in ER graphs, giving rise to the crossover for $\langle N_V\rangle$ shown in panels (c) and (d).

It is quite remarkable that for large $\lambda$ values and for $c=1.0$ [panels (b) and (d)] the number of vaccinated individuals vanishes and the values of $\langle R\rangle$ \alessio{goes close to one} in a similar way as in the original network (without vaccination). Obviously, as the vaccine cost $c$ increases, the solution $\langle R\rangle\simeq 1$ spans across a larger interval of $\lambda$ values so that for large enough $c$ there is no vaccinated individual in the population and one finally recovers the typical $\langle R\rangle(\lambda)$ diagram of Fig. \ref{fig2}.(a).

\section{Conclusions}

In this work we have analyzed the evolution of voluntary vaccination in networked populations. At variance with classical approaches we have considered an evolutionary framework so that individuals facing the vaccination dilemma do not take the most rational strategy by considering the benefits associated  to each choice. On the contrary,  they are considered as replicating agents that imitate the strategies based on their previous experience. To this aim we have entangled the spreading dynamics of an influenza-like disease with an evolutionary framework taking place at the end of each season. Our results show that when vaccine is perfect (so that vaccinated individuals do not get infected) scale-free networks enhance both the vaccination behavior and the effective immunization of the population as compared with random graphs with homogeneous connectivity patterns.

By considering vaccine imperfection we obtain two remarkable results. First, we have shown that, for scale-free networks and low vaccine costs, there is a threshold value for the vaccine imperfection so that, for values lower than this threshold, vaccination behavior spans across the population and it is possible to suppress the disease for all the infection probabilities. \alessio{Instead}, when vaccine imperfection becomes large, agents are less prone to take it and the disease takes advantage of this risky behavior to spread \alessio{more} efficiently across the population. 

The other interesting result concerns the comparison between scale-free and homogeneous networks. We have shown that when imperfection appears the better performance of scale-free network is broken and there is a cross-over effect so that the number of infected (vaccinated) individuals increases (decreases) with respect to homogeneous networks when $\lambda$ is large enough. This cross-over results from the competition of two well-known dynamical properties of scale-free networks: the fast propagation of diseases and the promotion of cooperative behaviors. Thus, the ability of scale-free networks in promoting cooperative behaviors (here represented as paying the cost of taking vaccine) is threatened when payoffs are dependent on a related dynamical process (here the spreading of a disease) whose evolution is also affected (here enhanced) by the heterogeneity of the network.

\begin{acknowledgments}
J.G.G. acknowledges the hospitality of UPTC and useful discussions with S.L. Dorado and L.M. Flor\'{\i}a. This work has been partially supported by the Spanish MINECO under projects FIS2011-25167 and FIS2012-38266-C02-01, by the Comunidad de Arag\'on (Grupo FENOL) and the UPTC (Proyecto Capital Semilla). J.G.G. is supported by MINECO through the Ram\'on y Cajal program.
\end{acknowledgments}


\begin{thebibliography}{99}

%Networks general
\bibitem{rev:albert} R. Albert, and A.-L. Barab\'asi,  Rev. Mod. Phys. {\bf 74}, 47 (2002).

\bibitem{rev:newman} M.E.J. Newman, SIAM Rev. {\bf 45}, 167 (2003).

\bibitem{rev:bocc} S. Boccaletti, V. Latora, Y. Moreno, M. Chavez and, D.-U. Hwang,  Phys. Rep. {\bf 424}, 175 (2006).

%Epidemic Modeling
\bibitem{book1} D.J. Daley and J. Gani, {\em Epidemic Modeling} (Cambridge Univ Press, Cambridge, 1999).

\bibitem{book2} R.M. Anderson, R.M. May, and B. Anderson, {\em Infectious Diseases of Humans: Dynamics
and Control} (Oxford Univ Press, Oxford, 1992).

\bibitem{rev:doro} S.N. Dorogovtsev, A.V. Goltsev, and J.F.F. Mendes, Rev Mod. Phys. {\bf 80}, 1275 (2008).

%Epidemics on networks (general)
\bibitem{vespi} R. Pastor-Satorras and A. Vespignani, Phys. Rev. Lett. {\bf 86}, 3200 (2001).

\bibitem{vespi2} R. Pastor-Satorras and A. Vespignani, Phys. Rev. E {\bf 63}, 066117 (2001).

\bibitem{may} A. L. Lloyd, and R. M. May, Science {\bf 292}, 1316 (2001).

\bibitem{moreno} Y. Moreno, R. Pastor-Satorras, and A. Vespignani, Eur. Phys. J. B {\bf 26}, 521 (2002).

\bibitem{marc} M. Barth\'el\'emy, A. Barrat, R. Pastor-Satorras, and A.Vespignani, Phys. Rev. Lett. {\bf 92}, 178701 (2004).

\bibitem{gomez2} S. G\'omez, J. G\'omez-Garde\~nes, Y. Moreno, and A. Arenas, Phys. Rev. E {\bf 84}, 036105 (2011).

%Human contacts
\bibitem{sex1} F. Liljeros, C. R. Edling, L. A. N. Amaral, H. E. Stanley, and Y. Aberg, Nature {\bf 411}, 907 (2001).

\bibitem{amaral-2000} L. Amaral, A. Scala, M. Barth\'el\'emy, and H. E. Stanley, Proc. Natl. Acad. Sci. USA {\bf 97}, 11149 (2000).

\bibitem{sex2} J.M. Read and M.J. Keeling, Proc. R. Soc. London B {\bf 270}, 699 (2002).

\bibitem{sex3} J.M. Read and M.J. Keeling, Theo. Pop. Biol. {\bf 70}, 201 (2006).

\bibitem{sex4} K.T.D Eames and M.J. Keeling, Proc. Natl. Acad. Sci. USA {\bf 99}, 13330 (2002).

\bibitem{sex5} J. G\'omez-Garde\~nes, V. Latora, Y. Moreno, and E. Profumo, Proc. Natl. Acad. Sci. USA {\bf 105}, 1399 (2008).

%Adaptive epidemies

\bibitem{Gross} T. Gross, C. Dommar D'Lima, and B. Blasius, Phys. Rev. Lett. {\bf 96}, 208701 (2006).

\bibitem{Guerra} B. Guerra and J. G\'omez-Garde\~nes, Phys. Rev. E {\bf 82}, 035101(R) (2010).

%Immunization
\bibitem{imm1} R. Pastor-Satorras and A. Vespignani, Phys. Rev. E {\bf 65}, 036104 (2002).

\bibitem{imm2} R. Cohen, S. Havlin, and D. ben Avraham, Phys. Rev. Lett. {\bf 91} 247901 (2003).

\bibitem{cover}  P. Echenique, J. G\'omez-Garde\~nes, and Y. Moreno, Phys. Rev. E {\bf 71}, 035102 (2005). 

\bibitem{imm3} J. G\'omez-Garde\~nes, P. Echenique, and Y. Moreno, Eur. Phys. J. B {\bf 49}, 259 (2006).

%Previous works

\bibitem{bauch1} C.T. Bauch, A.P. Galvani, and D.J.D. Earn, Proc. Natl. Acad. Sci. USA {\bf 100}, 10564 (2003).

\bibitem{bauch4} C.T. Bauch, and D.J.D. Earn, Proc. Natl. Acad. Sci. USA {\bf 101}, 13391 (2004).

\bibitem{vardavas}  R. Vardavas, R. Breban, and S. Blower, Phys. Rev. E {\bf 76}, 031127 (2007).

\bibitem{vardavas2}  R. Breban, R. Vardavas, and S. Blower, PLoS Comput. Biol. {\bf 3}, e85 (2007).

\bibitem{bauch2} A. Perisic and C.T. Bauch, PLoS Comput. Biol. {\bf 5}, e1000280 (2008).

\bibitem{bauch3} D.M. Cornforth, T.C. Reluga, E. Shim, C.T. Bauch, A.P. Galvani, and L.A. Meyers, PLoS Comput. Biol. {\bf 7}, e1001062 (2011).

\bibitem{NJP} H. Zhang, J. Zhang, C. Zhou, M. Small, and B. Wang, New J. Phys. {\bf 12}, 023015 (2010).

%Games on networks and lattices


%bhattarcharyya,bauch2012,donofrio2012

\bibitem{bhattarcharyya} S. Bhattacharyya, and C.T. Bauch, J. Theor. Biol {\bf 257}, 276 (2010).

\bibitem{bauch2012} C.T. Bauch, and S. Bhattacharyya, PLoS Comput. Biol. {\bf 8} e1002452 (2012).

\bibitem{donofrio2012} A. D'Onofrio, P. Manfredi, and P. Poletti, PLoS One {\bf 7}, e45653 (2012).

\bibitem{gintis} H. Gintis, {\em Game Theory Evolving} (Princeton, NJ: Princeton University Press, 2000).

\bibitem{nowak} M.A. Nowak, and R.M. May, Nature. {\bf 359}, 826 (1992).  

\bibitem{wells2013} C.R. Wells, E.Y. Klein, and C.T. Bauch, PLoS Comput. Biol. {\bf 9}, e1002945 (2013).

\bibitem{prsb} F. Fu, D.I. Rosenbloom, L. Wang, and M.A. Nowak, Proc. Roy. Soc. B {\bf278}, 42 (2011). 
%Imitation dynamics of vaccination behaviour on social networks

\bibitem{szabo} G. Szab\'o and G. Fath, Phys. Rep. {\bf 446}, 97 (2007).

\bibitem{anxorev} C.P. Roca, J.A. Cuesta, and A. S\'anchez, Phys. Life Rev. {\bf 6}, 208 (2009). 

\bibitem{our} M. Perc, J. G\'omez-Garde\~nes, A. Szolnoki, L.M. Flor\'{\i}a, and Y. Moreno, J. Roy. Soc. Interface {\bf 10}, 20120997 (2013).



\bibitem{santos} F.C. Santos and J.M. Pacheco,  Phys. Rev. Lett. {\bf 95}, 098104 (2005).

\bibitem{campillo} J. G\'omez-Garde\~nes, M. Campillo, L.M. Flor\'{\i}a, and Y. Moreno, Phys. Rev. Lett. {\bf 98}, 108103 (2007).

\bibitem{ER} P. Erd\H{o}s and A. R\'enyi, Publ. Math. Inst. Hung. Acad. Sci. \textbf{5}, 17 (1960).

\bibitem{BA} A.L. Barab\'asi, and R. Albert, Science {\bf 286}, 509 (1999).

%Fermi rule

\bibitem{fermi1} L.E. Blume, Games Econ. Behav. {\bf 5}, 387 (1993).

\bibitem{fermi2} G. Szab\'o and C. T\"oke, Phys. Rev. E {\bf 58}, 69 (1998).

\end{thebibliography}
\end{document}